\documentstyle[12pt,epsf]{article}

\newcommand{\beq}{\begin{equation}}
\newcommand{\eeq}{\end{equation}}
\newcommand{\beqa}{\begin{eqnarray}}
\newcommand{\eeqa}{\end{eqnarray}}
\newcommand{\no}{\nonumber}

\newcommand{\qq}{\qquad}
\newcommand{\mnod}{\stackrel{\circ}{M}} 

\newcommand{\tr}{\mbox{tr}}

\begin{document}

\hfill 

\hfill 

\bigskip\bigskip

\begin{center}

{{\Large\bf Resonances in weak nonleptonic $\Omega^-$ decay}}

\end{center}

\vspace{.4in}

\begin{center}
{\large Bu\={g}ra Borasoy\footnote{email: 
         borasoy@physik.tu-muenchen.de}$^\dagger$
 and Barry R. Holstein\footnote{email: holstein@phast.umass.edu }$^\#$}

\bigskip

\bigskip

$^\dagger$Physik Department\\
Technische Universit{\"a}t M{\"u}nchen\\
D-85747 Garching, Germany 

\bigskip

$^\#$Department of Physics and Astronomy\\
University of Massachusetts\\
Amherst, MA 01003, USA\\

\vspace{.2in}

\end{center}

\vspace{.7in}

\thispagestyle{empty} 

\begin{abstract}
We examine the importance of $J^P = \frac{1}{2}^+,\frac{1}{2}^-$ 
resonances for weak nonleptonic 
$\Omega^-$ decays within the framework of chiral perturbation theory.
The spin-1/2 resonances are included into an effective theory and
tree contributions to the $\Omega^-$ decays are calculated. We find
significant contributions to the decay amplitudes and satisfactory
agreement with experiment. This confirms and extends previous results
wherein such spin-1/2 resonances were included in nonleptonic 
and radiative-nonleptonic hyperon decays.
\end{abstract}

\vfill

\section{Introduction}  

Among the $J^P = \frac{3}{2}^+$ decuplet states only the 
$\Omega^-$ decays weakly into 
nonleptonic channels. The allowed two-body decay modes to the lowest-lying
baryon octet are $\Omega^- \rightarrow \Lambda K^-, 
\Omega^- \rightarrow \Xi^0 \pi^-$ and $\Omega^- \rightarrow \Xi^- \pi^0$.
The pertinent decay amplitudes have been calculated within the framework
of heavy baryon chiral pertubation theory at tree level \cite{Jen}
and also at one-loop order including the lowest nonanalytic 
corrections.\cite{EMS}
Within these investigations the authors come to the conclusion that the
decay amplitudes are well described by values for the weak
parameters of the Lagrangian determined from a fit to the s-wave
amplitudes of the nonleptonic decays of the octet baryons.
But the same values inserted into the expressions for the p-wave
amplitudes of the nonleptonic hyperon decays do not 
allow a good description
of the experimental data.\cite{Jen} 
Therefore, it is questionable if the chosen
values for the weak parameters are appropriate.

An intriguing  approach for nonleptonic
hyperon decays was examined by Le Yaouanc et al., who assert 
that a reasonable fit for both s- {\it and} p-waves can be provided
by appending pole contributions from $SU(6)$ (70,$1^-$) states
to the s-wave amplitudes.\cite{LeY} 
Their calculations were performed in a simple
constituent quark model and appeared to be able to provide a resolution of the 
s- and p-wave dilemma.
In a previous work \cite{BH1} we have studied this approach within a chiral
framework and have shown that it is indeed possible to find a simultaneous
fit to both s- and p-wave hyperon decay amplitudes if contributions
from the lowest lying 1/2$^-$ and 1/2$^+$ baryon octet resonant states
are included in the formalism.

Other weak processes involving hyperons are, 
e.g., the radiative-nonleptonic hyperon decays. 
The primary problem of radiative hyperon decay has been to understand
the large negative value found experimentally for
the asymmetry parameter in polarized $\Sigma^+ \rightarrow
p \gamma$ decay.\cite{pdg} 
The difficulty here is associated with the restrictions
posed by Hara's theorem, which requires the vanishing of this asymmetry
in the $SU(3)$ limit.\cite{Har}
Recent work involving the calculation of chiral loops has also not lead to
a resolution, although slightly larger asymmetries can be 
accomodated.\cite{Neu}
In \cite{BH2} we examined the contribution of 1/2$^-$ and 
1/2$^+$ baryon intermediate states to radiative hyperon decay
within the framework of chiral perturbation theory.
We obtained reasonable predictions for the decay amplitudes and a
significant negative value for the $\Sigma^+ \rightarrow
p \gamma$ asymmetry as a very natural result of this picture, even
though Hara's theorem is satisfied.
Thus, the inclusion of spin-1/2 resonances 
provides a reasonable explanation of the importance of higher order 
counterterms and gives a satisfactory picture of both radiative and
nonradiative nonleptonic hyperon decays.

In the present paper we extend the discussion to weak nonleptonic
$\Omega^-$ transitions and investigate the validity of this approach for
these decays.
In the next Section we introduce the effective weak and 
strong Lagrangian including resonant states and 
evaluate the pole diagram contributions of both ground state baryons
and resonant states to nonleptonic $\Omega^-$ decay.
Numerical results are presented in Sec. 3, and in Sec. 4 we conclude with
a short summary. In the Appendix, we determine the strong couplings
of the decuplet to the spin-1/2 resonances by a fit to the strong
decays of these spin-1/2 resonances.

\section{Weak nonleptonic $\Omega^-$ decay}

There are three two-body decay modes of the $\Omega^-$ to the ground state
baryon octet:
$\Omega^- \rightarrow \Lambda K^-, 
\Omega^- \rightarrow \Xi^0 \pi^-$ and $\Omega^- \rightarrow \Xi^- \pi^0$.
Phenomenologically, the matrix elements of these decays can each be expressed
in terms of a parity-conserving p-wave amplitude $A_{ij}^{(P)}$
and a parity-violating d-wave amplitude $A_{ij}^{(D)}$
\beq
{\cal A}( \Omega^- \rightarrow B_i \, \phi_j) =
\bar{u}_{B_i} \Big\{ \, A_{ij}^{(P)} q_\mu + 
       \, A_{ij}^{(D)}\gamma_5 q_\mu \Big\} u_{\Omega^-}^\mu ,
\eeq
where $B_i$ and $\phi_j$ are the ground state baryon and 
Goldstone boson, respectively,
and $q_\mu$ is the four-momentum of the outgoing kaon or pion.
The underlying strangeness-changing Hamiltonian
transforms under $SU(3)_L \times SU(3)_R$ as $(8_L, 1_R) \oplus (27_L,1_R)$
and, experimentally, the octet piece dominates over the 27-plet by a factor
of twenty or so in nonleptonic hyperon decay. 
In the case of $\Omega^-$ decay the corresponding octet
dominance contribution implies the relation
\beq
{\cal A}(\Omega^- \rightarrow \Xi^0 \pi^-) -
\sqrt{2} {\cal A}(\Omega^- \rightarrow \Xi^- \pi^0) =0 
\eeq
which holds both for p- and d-waves. The experimentally measured
ratio of the decay widths --- $\Gamma(\Omega^- \rightarrow \Xi^0 \pi^-)/
\Gamma(\Omega^- \rightarrow \Xi^- \pi^0) = 2.65 \pm 0.20$ --- is
in disagreement with the isospin prediction of 2. 
This indicates that the $\Delta I = 3/2$ amplitude in
$\Gamma(\Omega^- \rightarrow \Xi \pi)$ may be significantly
larger than in nonleptonic hyperon decay \cite{TV} and 
could signal a violation of the $\Delta I = 1/2$ rule 
for the $\Omega^-$ decay.\cite{CG}
We do not study this issue here and have, therefore, neglected
the 27-plet contribution.
We prefer not to work in the isospin limit, however, and
do not use this isospin relation to eliminate one of the decay
amplitudes but rather attempt to find a simultaneous fit to the three
decay modes of the $\Omega^-$.

The purpose of this work is to study the role of spin-1/2 resonances
in $\Omega^-$ decays. To this end, it is sufficient to work at tree
level. This is also necessary because the calculations \cite{BH1, BH2}
which provide values for the weak parameters involved in
these decays have been performed at tree level, so that 
it would be inconsistent
to use these values for the weak parameters in a loop calculation
for the $\Omega^-$ decays.
In the rest frame of the $\Omega^-$
the amplitudes are related to the decay width via
\beq    \label{width}
\Gamma = \frac{1}{12 \pi} \frac{|{\bf q}|}{M_\Omega} ( E_\phi^2- m_\phi^2)
\Big[ |A^{(P)}|^2 (E_B + M_B ) + |A^{(D)}|^2 (E_B - M_B ) \Big]
\eeq
with ${\bf q}$ being the three-momentum of the outgoing pseudoscalar.
In previous work the calculation of the d-wave amplitudes
$A^{(D)}$ has generally been neglected, since its contribution to the
decay width is suppressed by kinematical factors. However, in order
to evaluate the decay parameters on the other hand, one {\it must}
include the d-waves.
Thus, e.g., the asymmetry parameter is 
\beq
\alpha = \frac{2 \mbox{Re} ( A^{(P)*} \bar{A}^{(D)} )}{ |A^{(P)}|^2 
          + |\bar{A}^{(D)}|^2 }
\eeq
with $\bar{A}^{(D)} = [( E_B-M_B)/(E_B+M_B)]^{1/2} A^{(D)}$.\cite{pdg}

\subsection{The effective Lagrangian}
The effective Lagrangian can be decomposed into a strong and weak component
\beq
{\cal L}_{\mbox{eff}} = {\cal L}^{(S)} + {\cal L}^{(W)}.
\eeq
As mentioned above, since we will be using results from our previous work,
which considered the role of these resonances in nonleptonic hyperon
decay at tree level, we restrict ourselves to tree level
for the present study.
We first consider the Lagrangian in the absence of spin-1/2 resonances.
The strong part consists of the free kinetic Lagrangians
of the decuplet, the ground state baryon octet and the Goldstone bosons
together with an interaction term 
\beq
{\cal L}^{(S)} = {\cal L}_{kin} + {\cal L}_{\Delta B \phi} .
\eeq
The kinetic component reads
\beqa
{\cal L}_{kin} &=&  
    i \, \tr \big( \bar{B} \gamma_{\mu} [  D^{\mu} , B] \big) -
     \mnod \, \tr \big( \bar{B} B \big) +
    \frac{F_\pi^2}{4} \, {\rm tr} \big( u_{\mu} u^{\mu}\big)
  + \frac{F_\pi^2}{4} \, {\rm tr} \big( \chi_+ \big)  \no \\
 &+&  \bar{\Delta}^\alpha \bigg[ (-i \partial \!\!\!/ + M_\Delta) 
      g_{\alpha \beta} + i (\gamma_\alpha \partial_\beta +
     \gamma_\beta \partial_\alpha ) - i \gamma_\alpha \partial \!\!\!/
    \gamma_\beta - M_\Delta \gamma_\alpha \gamma_\beta  \bigg] \Delta^\beta
\eeqa
with $\mnod$ being the octet baryon mass in the chiral limit.
The pseudoscalar Goldstone fields ($\phi = \pi, K, \eta$) are collected in
the  $3 \times 3$ unimodular, unitary matrix $U(x)$, 
\begin{equation}
 U(\phi) = u^2 (\phi) = \exp \lbrace 2 i \phi / F_\pi \rbrace  \qq ,\qq
u_{\mu} = i u^\dagger \nabla_{\mu} U u^\dagger
\end{equation}
where $F_\pi \simeq 92.4$ MeV is the pion decay constant,
\begin{eqnarray}
 \phi =  \frac{1}{\sqrt{2}}  \left(
\matrix { {1\over \sqrt 2} \pi^0 + {1 \over \sqrt 6} \eta
&\pi^+ &K^+ \nonumber \\
\pi^-
        & -{1\over \sqrt 2} \pi^0 + {1 \over \sqrt 6} \eta & K^0
        \nonumber \\
K^-
        &  \bar{K^0}&- {2 \over \sqrt 6} \eta  \nonumber \\} 
\!\!\!\!\!\!\!\!\!\!\!\!\!\!\! \right) \, \, \, \, \,  
\end{eqnarray}
represents the contraction of the pseudoscalar fields with the Gell-Mann
matrices and
$B$ is the standard $SU(3)$ matrix representation of the low-lying
spin-1/2 baryons $( N, \Lambda, \Sigma, \Xi)$.
Here $\chi_+ = u^\dagger
\chi u^\dagger + u \chi^\dagger u$ is proportional to the quark mass
matrix ${\cal M}$  = ${\rm diag}(m_u,m_d,m_s)$ via $\chi = 2 B {\cal
  M}$. Also, $B = - \langle 0 | \bar{q} q | 0 \rangle / F_\pi^2$ is
the order parameter of the spontaneous symmetry violation,
and we assume $B \gg F_\pi$. 
The propagator of the spin-3/2 fields (denoted generically
by $\Delta$) is given by
\begin{equation}
G_{\beta \delta} (p) = -i\frac{p \!\!/ + M_\Delta}{p^2-M_\Delta^2} 
\, \biggl( g_{\beta \delta} - \frac{1}{3} \gamma_\beta \gamma_\delta -
\frac{2 p_\beta p_\delta}{3M_\Delta^2} + \frac{p_\beta \gamma_\delta -
p_\delta \gamma_\beta}{3 M_\Delta} \, \biggr) \, \, \, ,
\end{equation}
with $M_\Delta = 1.38$~GeV being the average decuplet mass.

The interaction Lagrangian between the spin-3/2
fields, the baryon octet and the Goldstone bosons reads
\begin{equation}
{\cal L}_{\Delta B \phi} = 
\frac{{\cal C}}{2} \, \biggl\{ \bar{\Delta}^{\mu ,abc} \, 
 \Theta_{\mu \nu} (Z) \, (u^\nu)_a^i \, B_b^j \, 
\epsilon_{cij}- \bar{B}^b_i \, (u^\nu)^a_j \,  \Theta_{\nu \mu} (Z) \,
{\Delta}^\mu_{abc} \,\epsilon^{cij}\, \biggr\} \, \,  \, ,
\end{equation}
where $a, b, \ldots , j$ are SU(3)$_f$ indices
and the coupling constant $1.2 < {\cal C} < 1.8 $ has been
determined from the decays $\Delta \to
B \pi$.\cite{JM} In our application, we use the mean value $C=1.5$.
The Dirac matrix operator $\Theta_{\mu \nu} (Z)$ is given in general by
\begin{equation}
\Theta_{\mu \nu} (Z) = g_{\mu \nu} - \biggl(Z + \frac{1}{2} \biggr) \,
\gamma_\mu \, \gamma_\nu \, \, \, \, .
\label{theta}
\end{equation}
However, the off-shell parameter $Z$ does not contribute at tree level
because of the subsidiary condition $\gamma_\mu u^\mu_{\Omega^-} =0$.

The weak Lagrangian can be written as
\beq  \label{hpi}
{\cal L}^{(W)}  =
         d \, \tr \Big( \bar{B}  \{ h_+ , B\} \Big) + 
         f \, \tr \Big( \bar{B}  [ h_+ , B ] \Big)  +
         \frac{F_\pi^2}{4} \, h_{\pi} \, {\rm tr}
        \Big( h_+ u_{\mu} u^{\mu}\Big)   
     + \, h_c \, \bar{\Delta}^{\mu, abc} (h_+)_a^d \Delta_{\mu, dbc},
\eeq
where we have defined
\beq
h_+ = u^{\dagger} h u +  u^{\dagger} h^{\dagger} u  \qquad , 
\eeq
with
$h^{a}_{b} = \delta^{a}_{2} \delta^{3}_{b}$ being
the weak transition matrix. 
(Note that $h_+$   transforms as a matter field.) 
In Eq.(\ref{hpi})  the weak coupling $h_{\pi}$ is known from
weak nonleptonic kaon decays --- $h_{\pi} = 3.2 \times 10^{-7}$ --- 
while in our previous work the LECs $d$ and $f$ have been determined from 
nonleptonic hyperon decay: $d=0.44 \times 10^{-7}$ GeV, 
$f=-0.50 \times 10^{-7}$ GeV.
In \cite{BH1} 
we included lowest lying spin-1/2$^+$ and 1/2$^-$ resonances
in the theory and performed a tree level calculation.
Integrating out the heavy degrees of freedom provides
then a plausible estimate of the weak counterterms and
a satisfactory fit for both s- and p-waves was achieved.
Since we herein apply this scheme for the related weak $\Omega^-$ decays 
we will use the values for $d$ and $f$ from \cite{BH1}.
The parameter $h_c$ does not appear in the tree-level calculation of the
nonleptonic hyperon decays in \cite{BH1,BH2}. 
We therefore consider it as a free parameter and determine its value
from a fit to the weak $\Omega^-$ decays.

We now
proceed to append the contribution from spin-1/2 resonances.
In \cite{LeY,LeY1} it was argued that in a simple constituent 
quark model inclusion of the
lowest lying spin 1/2$^-$ octet from the (70,1$^-$) multiplet leads
to significant improvements in both radiative and nonleptonic hyperon 
decays, and we confirmed in two recent calculations \cite{BH1,BH2} 
that there indeed exist
significant contributions from such resonances to the nonleptonic hyperon
decays in the framework of chiral perturbation theory. 
We begin therefore with the inclusion of 
the octet of spin-parity 1/2$^-$ states, which include the
well-established 
states $N(1535)$ and $\Lambda(1405)$.  As for the rest
of the predicted 1/2$^-$ states there are a number of not so well-established
states in the same mass range --- cf. \cite{LeY} and references therein.
In order to include such resonances one begins by writing down the most
general Lagrangian at lowest order which exhibits the same symmetries
as the underlying theory, i.e. Lorentz invariance and chiral symmetry.
For the strong part we require invariance under $C$ and $P$ transformations
separately, while
the weak piece is invariant under $CPS$ transformations,
where the transformation $S$
interchanges down and strange quarks in the Lagrangian.
We will work in the $CP$-conserving limit so that all LECs are real,
and denote the 1/2$^-$ octet by $R$. 
Another important multiplet of excited states 
is the octet of
Roper-like spin-1/2$^+$ fields. While it was argued in \cite{JM} that
these play no significant role, a more recent study seems to indicate that one
cannot neglect contributions from such states to,
e.g., decuplet magnetic moments.\cite{BaM} 
Also in \cite{BH1,BH2} we found that inclusion of these states was essential
both in nonleptonic and radiative hyperon decay to achieve 
experimental agreement.
It is thus
important to include also the contribution of these baryon
resonances here.
The Roper octet, which we denote by $B^*$, 
consists of the $N^*(1440)$, the $\Sigma^*(1660)$, the 
$\Lambda^*(1600)$ and the $\Xi^*(1620?)$.

The resonance kinetic term is straightforward
\beq
{\cal L}_{kin}  
 = 
 i \, \tr \Big( \bar{R} \gamma_{\mu} [  D^{\mu} , R] \Big) -
M_R \, \tr \Big( \bar{R} R \Big)
+i \, \tr \Big( \bar{B}^* \gamma_{\mu} [  D^{\mu} , B^*] \Big) -
M_{B^*} \, \tr \Big( \bar{B}^* B^* \Big)
\eeq
with $M_R$ and $M_{B^*}$ being the masses of the resonance octets 
in the chiral limit.
The strong interaction Lagrangian relevant to 
the processes considered here reads
\beqa
{\cal L}_{\Delta \phi B^*/R} &=& 
\frac{{\cal C_{B^*}}}{2} \, \biggl\{ \bar{\Delta}^{\mu ,abc} \, 
 \Theta_{\mu \nu} (Z) \, (u^\nu)_a^i \, B_b^{*j} \, 
\epsilon_{cij}- \bar{B}^{*b}_i \, (u^\nu)^a_j \,  \Theta_{\nu \mu} (Z) \,
{\Delta}^\mu_{abc} \,\epsilon^{cij}\, \biggr\}  \no \\
&+&  i \frac{s_c}{2} \, \biggl\{ \bar{\Delta}^{\mu ,abc} \, 
 \Theta_{\mu \nu} (Z) \, (u^\nu)_a^i \, \gamma_5 \, R_b^j \, 
\epsilon_{cij}- \bar{R}^b_i \, (u^\nu)^a_j \,  \Theta_{\nu \mu} (Z) \,
\gamma_5 \, {\Delta}^\mu_{abc} \,\epsilon^{cij}\, \biggr\}  \no \\
\eeqa
and the couplings $C_{B^*}$ and $s_c$ can be determined from a fit
to the strong decays of the spin-1/2 resonances to 
$\Delta(1232)$ --- cf. App. A.
The corresponding weak Lagrangian is
\beqa
{\cal L}^{W}  &=& 
d^* \Big[ \, \tr \Big( \bar{B}^*  \{h_+ , B\} \Big)
       + \tr \Big( \bar{B}  \{ h_+, B^* \} \Big) \: \Big] +
 f^* \Big[ \, \tr \Big( \bar{B}^*  [h_+ , B] \Big)
       + \tr \Big( \bar{B}  [ h_+, B^*] \Big) \: \Big] \no \\
&+ & i w_d \Big[ \, \tr \Big( \bar{R}  \{h_+ , B\} \Big)
       - \tr \Big( \bar{B}  \{ h_+, R\} \Big) \: \Big] +
  i w_f \Big[ \, \tr \Big( \bar{R}  [h_+ , B] \Big)
       - \tr \Big( \bar{B}  [ h_+, R] \Big) \: \Big] 
\eeqa
with four couplings $d^*, f^*$ and $w_d, w_f$ which have been determined from
a fit to the nonleptonic hyperon decays in \cite{BH1}.

There is thus only a single unknown parameter
in this approach --- $h_c$ --- once the weak 
couplings are fixed from nonleptonic hyperon decays. 
The inclusion of the spin-1/2 resonant states does not lead to any new
unknown parameters, so that 
study of the weak nonleptonic $\Omega^-$ decays provides
a nontrivial check on the role of spin-1/2 resonances
in weak baryon decays --- extending this approach to the case of
weak decuplet decays.

\subsection{Pole contributions}
In this section we consider the tree diagrams which contribute to weak
$\Omega^-$ decay. The graphs involving only the decuplet and the ground
state baryon octet are depicted in Fig. 1.  
Graphs 1a and 1b contribute
to the decay $\Omega^- \rightarrow \Lambda K^-$, while
graphs 1b and 1c deliver contributions to the pionic decays
$\Omega^- \rightarrow \Xi \pi$. (Note that diagram 1c with the
weak decay of the Goldstone boson has been
neglected in \cite{Jen} and \cite{EMS}.)
The diagrams in Fig. 1 contribute only to the parity-conserving p-wave 
amplitudes $A^{(P)}$, yielding
\beqa
A^{(P)}_{\Lambda K}  & = &
 \frac{C}{2 \sqrt{3} F_\pi} \left( \frac{d-3f}{M_\Lambda-M_{\Xi^0}}
  + \frac{h_c}{M_\Omega-M_{\Xi^*}} \right) \no \\
A^{(P)}_{\Xi^0 \pi^-}  & = &
 \frac{C}{\sqrt{2} F_\pi} \left( \frac{h_c}{3(M_\Omega-M_{\Xi^*})}
  + \frac{h_\pi m^2_{\pi^-}}{2(m^2_{K^-}-m^2_{\pi^-})} \right) \no \\
A^{(P)}_{\Xi^- \pi^0}  & = &
 \frac{C}{2 F_\pi} \left( \frac{h_c}{3(M_\Omega-M_{\Xi^*})}
  + \frac{h_\pi m^2_{\pi^0}}{2(m^2_{K^0}-m^2_{\pi^0})} \right) .
\eeqa
Here, we have used the physical meson masses and have 
replaced the baryon masses in the chiral limit by their
physical values, which is consistent to the order we are working.
There are no contributions from these diagrams to the parity-violating
d-wave amplitudes $A^{(D)}$, so that at this order
there is in each case a vanishing
asymmetry parameter $\alpha$.

We now include the spin-1/2 resonances,
which contribute only to the decay $\Omega^- \rightarrow  \Lambda K^-$ 
through the diagram depicted in Fig. 2. 
One obtains
the result
\beqa
A^{(P)}_{\Lambda K}  & = &
 \frac{C_{B^*}}{2 \sqrt{3} F_\pi} \frac{d^*-3f^*}{M_\Lambda-M_{B^*}} \no \\
A^{(D)}_{\Lambda K}  & = &
 \frac{s_c}{2 \sqrt{3} F_\pi} \frac{w_d-3w_f}{M_\Lambda-M_R}  
\eeqa
and there are no further contributions from the spin-1/2 resonant states.

\section{Numerical results}
In this section we present the results obtained by a 
fit to the three weak $\Omega^-$ decays. 
(Note that when applying Eq.(\ref{width}) in order to perform the fit,
we use the physical values for the masses of the outgoing particles.
Therefore, we obtain a ratio  $\Gamma(\Omega^- \rightarrow \Xi^0 \pi^-)/
\Gamma(\Omega^- \rightarrow \Xi^- \pi^0)$
which is slightly different from the isospin prediction of 2.
However, the 30\% effect found experimentally is outside
our approach.)
Our fit including the spin-1/2 resonances
leads to $h_c = 0.39 \times 10^{-7}$ GeV and in units of $10^{-15}$ GeV
\beqa
\Gamma(\Omega^- \rightarrow \Lambda K^-) & = &  5.61  \qq (5.42 \pm 0.06)\no \\
\Gamma(\Omega^- \rightarrow \Xi^0 \pi^-) & = &  1.62 \qq (1.89\pm 0.06)\no \\
\Gamma(\Omega^- \rightarrow \Xi^- \pi^0) & = &  0.76  \qq (0.69 \pm 0.03) , 
\eeqa
where we have given the experimental numbers in brackets. (Note that the ratio
$\Gamma(\Omega^- \rightarrow \Xi^0 \pi^-)/
\Gamma(\Omega^- \rightarrow \Xi^- \pi^0)$ is approximately 
2.1 in our approach.)
The inclusion of the spin-1/2 resonances predicts a finite decay parameter
$\alpha$ for the decay $\Omega^- \rightarrow \Lambda K^-$
\beq
\alpha_{\Lambda K} =  -0.015
\eeq
to be compared with an experimental value of $-0.026 \pm 0.026$,
while we obtain
vanishing asymmetry parameters for  the two pionic decay modes 
which are also consistent with the experimental values
$\alpha_{\Xi^0 \pi^-} = 0.09 \pm 0.14$ and
$\alpha_{\Xi^0 \pi^-} = 0.05 \pm 0.21$.
\footnote{The results
from a fit without the inclusion of the spin-1/2
resonances read in units of $10^{-15}$ GeV
$\Gamma(\Omega^- \rightarrow \Lambda K)   =   4.64,
\Gamma(\Omega^- \rightarrow \Xi^0 \pi^-)   =   0.69,
\Gamma(\Omega^- \rightarrow \Xi^- \pi^0)   =   0.32   $
with $h_c = 0.24 \times 10^{-7}$ GeV
and vanishing asymmetry parameters in each case.
Interestingly, we do not have a satisfactory fit to experiment. However,
it is not really fair to compare these results 
to the fit including the spin-1/2 resonances, since
we employ values for $d$ and $f$ which
have been determined in \cite{BH1} from a fit to the nonleptonic
decays after the inclusion of such states.}

Our results are only indicative. A full discussion would have to include both
the effects of chiral loops as well as contributions from higher resonances.
In this regard, we do not anticipate our results to be able to precisely
reproduce the experimental values for the decay widths and the
asymmetry parameters.
Although it would be desirable to estimate the error of our tree level
result from higher order chiral loop corrections by e.g. calculating
some typical loop contributions, it is then inconsistent to use results from
our previous work which considered the role of resonances in nonleptonic 
hyperon decay at tree level \cite{BH1, BH2}. 
Therein we derived estimates on the resonance 
parameters at tree level which we also use in the present investigation.
The inclusion of loops might change the numerical values of these parameters
somewhat and, therefore, alter the contribution from resonances
both in nonleptonic hyperon decay and $\Omega^-$ decay. 
In order to give a reliable estimate on chiral corrections in $\Omega^-$ decay,
one has to consider---in addition to chiral loops for this process---also the
the change in the resonance contribution due to the inclusion of loops in
nonleptonic hyperon decay. This is clearly beyond the scope of the present
investigation.
Our purpose herein is to examine whether the inclusion of
spin-1/2 resonances, which worked well in both ordinary
and radiative nonleptonic hyperon decay, can be extended 
successfully to the case
of weak decuplet decays. Our results suggest that the spin-1/2 resonances
also play an essential role in weak nonleptonic decuplet decay.

Before leaving this section, it is interesting to note that while
our results are purely phenomenological, it is possible to compare the
fit value of $h_c$, which leads in our model
to $\langle \Xi^{*-} | H_W | \Omega^- \rangle =
0.22 \times 10^{-7}$ GeV, with what is expected based upon quark model
considerations.
A constituent quark model approach to the weak $\Xi^{*-} - \Omega^-$
matrix element would include two structures --- one from the direct
four-quark weak Hamiltonian and a second from the related penguin term.
However, the direct contribution to the $\Xi^{*-} - \Omega^-$
transition vanishes because of lack of multiple $s, \bar{s}$ quarks in $H_W$,
so that a non-zero value for this matrix element can only arise from
the penguin term. The size of this contribution is somewhat larger
than what one might expect based on model calculations
with perturbatively calculated Wilson coefficients evolved to 
$\mu \simeq 1$ GeV, 
but is consistent with the expected size of such effects
when scaled into the low energy region $\mu \simeq 0.2$ GeV.\cite{DGH}

\section{Summary}
In this work we have 
examined the role of spin-1/2 intermediate state resonances in weak 
nonleptonic $\Omega^-$ decay. This study also serves as a consistency check
for the results from nonleptonic hyperon decay.
In two recent papers we were able to show that these 
resonances play an essential role for both ordinary and radiative
nonleptonic hyperon decay.\cite{BH1,BH2}  A much improved agreement
with the experimental values
for s- and p-waves in ordinary hyperon decay is brought about
even at tree level and a significant negative
value for the asymmetry parameter in the radiative decay
$\Sigma^+ \rightarrow p \gamma$ is obtained
as a very natural result within this picture.

In the present work we have shown
that this approach can be extended to the case of weak decuplet decays.
The pertinent Lagrangian without spin-1/2 resonances 
has a single unknown weak parameter once the remaining weak parameters are
fixed by the nonleptonic hyperon decays. 
Inclusion of the spin-1/2 resonances does not lead to any additional
unknown parameters, since the two strong couplings involving
the decuplet can be fixed from the strong decays of the spin-1/2 resonances
to the decuplet.  
The spin-1/2 resonances contribute only to the decay
$\Omega^- \rightarrow \Lambda K^-$.
We determine the unknown parameter
by a least-squares fit to the three two body decay modes of the $\Omega^-$.
One obtains satisfactory agreement with experimental
data and a nonvanishing asymmetry parameter
$\alpha_{\Lambda K} = - 0.014$ which lies within the experimental range ---
$\alpha_{\Lambda K} = - 0.026 \pm 0.026$.
Since we do not work in the isospin limit we obtain a ratio
$\Gamma(\Omega^- \rightarrow \Xi^0 \pi^-)/
\Gamma(\Omega^- \rightarrow \Xi^- \pi^0) = 2.1$ which is different
from the isospin prediction of 2. This ratio is measured to be
approximately 2.65.\cite{pdg}
This indicates that the $\Delta I = 3/2$ amplitude from
the current measurement of the rates for 
$\Gamma(\Omega^- \rightarrow \Xi \pi)$ appears to be significantly
larger than in nonleptonic hyperon decay \cite{TV} and 
could signal a violation of the $\Delta I = 1/2$ rule 
for the $\Omega^-$ decay.\cite{CG}
We do not test that here and have, therefore, neglected
the 27-plet contribution.

Our study suggests that the approach of including spin-1/2 resonances in
nonleptonic hyperon decay can be extended to weak decuplet decays 
giving a satisfactory picture of weak $\Omega^-$ decay.
In order to make a more definite statement one should, of course,
go to higher orders and include meson loops as well as the contributions
from additional resonances. However, this is beyond the scope
of the present investigation.

\section*{Acknowledgements}
This work was supported in part by the Deutsche Forschungsgemeinschaft
and by the National Science Foundation.

\appendix 
\def\theequation{\Alph{section}.\arabic{equation}}
\setcounter{equation}{0}
\section{Determination of the spin-1/2 resonance couplings 
          with the decuplet} \label{app:a}

The decays listed in the particle data book,
which determine the coupling constants $C_{B^*}$ and $s_c$,
are $N(1440) \rightarrow \Delta(1232) \pi$ and
$N(1535) \rightarrow \Delta(1232) \pi$.
For the sake of brevity we will only present the case of the 
1/2$^+$-resonances. The parameter $s_c$ of the 1/2$^-$ octet has been
determined in a completely analogous way and we just present the result.
The width follows via
\beq
\Gamma = \frac{1}{8 \pi M_{B^*}^2} |{\bf k}_\pi| |{\cal T}|^2
\eeq
with
\beq 
|{\bf k}_\pi| = \frac{1}{2 M_{B^*}} \Big[ (M_{B^*}^2-(M_\Delta+m_\pi)^2)
                 \, (M_{B^*}^2-(M_\Delta-m_\pi)^2) \Big]^{1/2}
\eeq
being the three-momentum of the pion
in the rest frame of the spin-1/2 resonance.
The terms $M_{B^*}$ and $M_\Delta$ are the masses of the octet resonance 
and the decuplet, respectively.
For the transition matrix one obtains
\beq
|{\cal T}|^2 = \frac{C_{B^*}^2 M_{B^*}}{3 F_\pi^2 M^2_\Delta}  
    (E_\Delta + M_\Delta) 
  \Big[ (M_{B^*}^2 - M_\Delta^2 - m_\pi^2)^2 - 4 M_\Delta^2 m_\pi^2 \Big] .
\eeq
Using the experimental value for the decay width
we arrive at the central value
\beq
C_{B^*} = 1.35
\eeq
where we have chosen the sign of $C_{B^*}$ to be positive 
since this  leads to a better fit for the $\Omega^-$ decays.
From a similar calculation in the case of the 1/2$^-$-resonances
one obtains
\beq
s_c = -0.85 .
\eeq
We do not present the uncertainties in these parameters here, since
for the purpose of our considerations a rough estimate of these constants
is sufficient.

\newpage


\section*{Figure captions}

\begin{enumerate}

\item[Fig.1] Shown are the diagrams without spin-1/2 resonances
         in weak nonleptonic $\Omega^-$ decay.
         Solid and dashed lines denote ground state baryons and
         pseudoscalar mesons, respectively.  The double line is
         a decuplet state. The solid square represents
         a weak vertex and the solid circle denotes a strong vertex.

\item[Fig.2] Shown is the diagram with spin-1/2 resonances
         in the weak $\Omega^- \rightarrow \Lambda K^-$ decay.
         Solid and dashed lines denote the $\Lambda$ and the
         kaon, respectively.  The double line is
         the $\Omega^-$ and the thick line denotes the spin-1/2
         resonance. The solid square represents
         a weak vertex and the solid circle denotes a strong vertex.

\end{enumerate}

\newpage

\begin{center}
 
\begin{figure}[bth]
\centering
\centerline{
\epsfbox{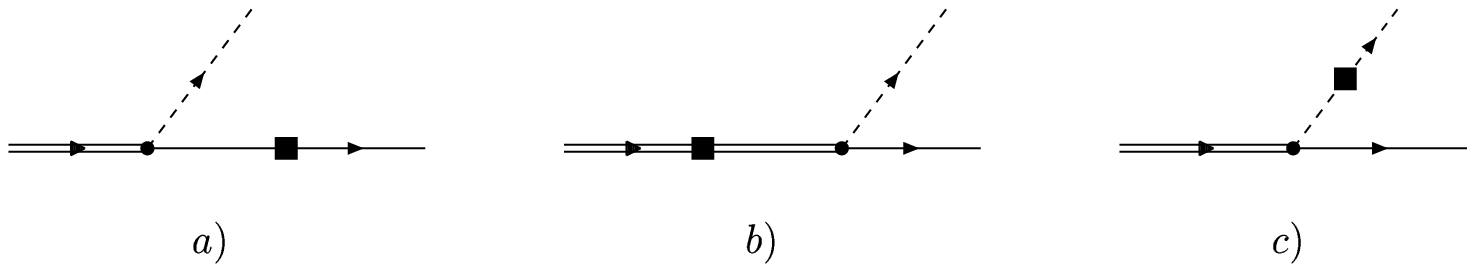}}
\end{figure}

\vskip 0.7cm

Figure 1

\vskip 1.5cm

\begin{figure}[tbh]
\centering
\centerline{
\epsfbox{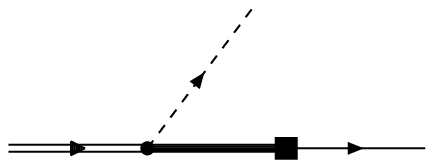}}
\end{figure}

\vskip 0.7cm

Figure 2

\end{center}


\begin{thebibliography}{99}

\bibitem{Jen}  E. Jenkins, Nucl. Phys. {\bf B375} (1992) 561 

\bibitem{EMS} D. Egolf, I. V. Melnikov, R. P. Springer, Phys. Lett.
               {\bf B451} (1999) 267

\bibitem{LeY} A. Le Yaouanc, O. Pene, J.-C. Raynal, L. Oliver,
               Nucl. Phys. {\bf B149} (1979) 321

\bibitem{BH1} B. Borasoy, B. R. Holstein, 
               Phys. Rev. {\bf D59} (1999) 094025

\bibitem{pdg} Particle Data Group, C. Caso et al., Eur. Phys. J. 
              {\bf C3} (1998) 1

\bibitem{Har} Y. Hara, Phys. Rev. Lett. {\bf 12} (1964) 378 

\bibitem{Neu} H. Neufeld, Nucl. Phys. {\bf B402} (1993) 166

\bibitem{BH2} B. Borasoy, B. R. Holstein, 
               Phys. Rev. {\bf D59} (1999) 054017

\bibitem{TV}  J. Tandean, G. Valencia, hep-ph/9810201

\bibitem{CG}  C. Carone, H. Georgi, Nucl. Phys. {\bf B375} (1992) 243

\bibitem{JM} E. Jenkins, A.V. Manohar, in ``Effective Field Theories
of the Standard Model'', ed. Ulf-G. Mei{\ss}ner, World Scientific,
Singapore, 1992

\bibitem{LeY1} M.B. Gavela, A. Le Yaouanc, L. Oliver, O. Pene, J.-C. Raynal,
             T.N. Pham, Phys. Lett. {\bf B101} (1981) 417

\bibitem{BaM} M.K. Banerjee, J. Milana, Phys. Rev. {\bf D54} (1996) 5804

\bibitem{DGH} J. F. Donoghue, E. Golowich, B. R. Holstein,
              Phys. Rep. {\bf 131} (1986) 319


\end{thebibliography}
\end{document}